\documentclass[twocolumn]{aastex701}
\usepackage{CJK}
\usepackage{bm}
\usepackage{{amsmath}}

\begin{document}
\begin{CJK*}{UTF8}{gbsn}

\title{Circumbinary disk formation through AGB star winds}

\author[0000-0002-7276-3694]{Shunquan Huang (黄顺权)}
\affiliation{Department of Physics and Astronomy, University of Nevada, Las Vegas, 4505 South Maryland Parkway, Las Vegas, NV 89154, USA}
\affiliation{Nevada Center for Astrophysics, University of Nevada, Las Vegas, 4505 South Maryland Parkway, Las Vegas, NV 89154, USA}
\email{huangs18@unlv.nevada.edu} 

\author[0000-0003-2401-7168]{Rebecca G. Martin}
\affiliation{Department of Physics and Astronomy, University of Nevada, Las Vegas, 4505 South Maryland Parkway, Las Vegas, NV 89154, USA}
\affiliation{Nevada Center for Astrophysics, University of Nevada, Las Vegas, 4505 South Maryland Parkway, Las Vegas, NV 89154, USA}
\email{rebecca.martin@unlv.edu} 

\begin{abstract}

Circumbinary disks are commonly observed around post-asymptotic giant branch (post-AGB) star binaries, yet their formation and especially their long-term evolution remain unclear.
We investigate this process using smoothed-particle hydrodynamics simulations of AGB star wind-binary interactions across different wind velocities, binary eccentricities, and mass ratios. 
When the wind is fast or the companion is relatively low-mass, corresponding to the Bondi-Hoyle regime, the outflow remains largely unbound and forms a spiral density pattern. 
In contrast, slower winds and more massive companions lead to wind Roche-lobe overflow (WRLOF), where a circumsingle disk forms around the companion and efficiently transfers angular momentum to the outflow, producing a circumbinary disk. 
We perform simulations for over $600$ binary orbital periods and find that the resulting disk properties depend sensitively on binary parameters, with higher eccentricities producing more extended and eccentric circumbinary disks, while lower mass companions reduce the disk density and growth rate. 
We further find that angular momentum transport within the circumbinary disk is dominated by spiral structures and shocks generated by the binary--wind interaction, corresponding to effective stresses comparable to or larger than the imposed viscosity.
These results show that wind-binary interactions can naturally generate diverse circumbinary disk morphologies observed in AGB and post-AGB star systems.

\end{abstract}



\section{Introduction} \label{sec:intro}
At the end of a star's life, it enters the asymptotic giant branch (AGB) phase if its mass is in the range $0.8-8\,\rm M_\odot$ (\citealt{2004agbs.book.....H}).
An AGB star can expand to a size significantly larger than that of its main-sequence (MS) companion and lose mass through a strong stellar wind (\citealt{2018A&ARv..26....1H}).
The AGB star wind is launched from the dust formation region around the star (\citealt{2005A&A...438..273V}).
When the gas escaping from the AGB star cools to the dust condensation temperature, dust grains form (\citealt{2022A&A...667A..75S,2024A&ARv..32....2S}).
These grains absorb the stellar radiation, heat up, and are accelerated outward, dragging the gas and driving the wind (\citealt{2020ApJ...892..110C,2023A&A...674A.122E}).

Stable Keplerian disks have been observationally detected around many post-AGB stars (\citealt{2006A&A...448..641D,2013A&A...557L..11B,2021A&A...648A..93G}).
All of these disks are circumbinary, and the post-AGB stars hosting them are found exclusively in binary systems (\citealt{2013A&A...557L..11B,2018Galax...6...97I}).
This evidence suggests that the formation of circumbinary disks is closely related to binarity.
These binaries have orbital periods ranging from about 200 to 3000 days, with most clustering around $\sim 1000$ days (\citealt{2018A&A...620A..85O,2018A&A...618A..58M,2019ApJ...877..110K,2022A&A...658A..36K}).

For binaries with short orbital periods, the companion may become engulfed by the expanded AGB star envelope, leading to a common-envelope (CE) phase (\citealt{1976IAUS...73...75P,2013A&ARv..21...59I}).
Such a phase has been proposed as a possible channel for the formation of circumbinary disks (\citealt{2011MNRAS.417.1466K,2016MNRAS.455.4351P,2016MNRAS.461.2527P}).
During the CE interaction, the envelope gains orbital energy and angular momentum from the binary through tidal dissipation and friction (\citealt{1993PASP..105.1373I}).
As a result, most of the envelope is ejected, potentially leaving behind a circumbinary disk (\citealt{2023MNRAS.521...35I,2019MNRAS.489..891H}).
For wider binaries, where the orbital separation exceeds the radius of the AGB star and a common envelope does not form, mass transfer between the two stars becomes important (\citealt{2012Sci...337..444S}).

The mode of mass transfer in binary systems depends strongly on the orbital separation.
When the AGB star expands to fill its Roche lobe without engulfing the companion, Roche-lobe overflow (RLOF) can occur (\citealt{1971ARA&A...9..183P,1983ApJ...268..368E,2014LRR....17....3P}).
If the binary separation is larger than the dust formation region, the wind can become fully accelerated (\citealt{2018A&ARv..26....1H,2020ApJ...892..110C}).
In this case, the companion can capture part of the wind through Bondi-Hoyle (BH) accretion (\citealt{1939PCPS...35..405H,1944MNRAS.104..273B,2004NewAR..48..843E}).
In the ideal BH accretion regime, the accretion rate is typically low, and the formation of a stable circumsecondary disk is difficult. This situation usually corresponds to binaries with relatively large separations.
In such cases, the wind is primarily gravitationally focused and scattered by the companion, producing a spiral density pattern in the outflow (\citealt{2020Sci...369.1497D,2023A&A...674A.122E,2024A&A...691A..57M}).

However, when the companion lies close to the dust formation region, it can interact with the wind before the outflow becomes fully accelerated, leading to wind Roche-lobe overflow (WRLOF) (\citealt{2007ASPC..372..397M,2013A&A...552A..26A,2019A&A...626A..68S,2024ApJ...969....8S}).
In this regime, although the AGB star itself does not fill its Roche lobe, the slow and dense outflow can effectively fill the Roche lobe of the donor.
The wind material can then be focused toward the companion through the inner Lagrange point (L1), forming a stream crossing L1 and potentially producing an accretion disk around the companion.
This process can significantly enhance the mass accretion rate onto the companion and increase the efficiency of angular momentum transfer between the wind and the binary system.

There have been many previous studies of circumbinary outflows in AGB star binary systems.
\cite{2016MNRAS.455.4351P,2016MNRAS.461.2527P} performed smoothed-particle hydrodynamics (SPH) simulations of a contact binary system undergoing a common-envelope phase and found outflows toward circumbinary orbits through the outer Lagrange point (L2) (see also \citealt{2019MNRAS.489..891H,2019MNRAS.484..631R}).
Circumbinary outflows through L2 have also been found in semi-detached (WRLOF) systems in many grid-based simulations (\citealt{2017MNRAS.468.4465C,2020ApJ...892..110C,2023MNRAS.519.1409L,2025ApJ...990..172S}).
These works provide important insights into mass transfer in binary systems and the formation of circumbinary outflows.
However, most of these simulations were run for less than $\sim100$ orbital periods, which limits their ability to study the long-term formation and evolution of circumbinary disks.
Therefore, the long-term evolution of circumbinary disks formed from AGB star winds remains poorly explored.

In this work, we perform hydrodynamical simulations of AGB star binary systems in both the BH accretion regime, where the wind is fully accelerated, and the wind Roche-lobe overflow (WRLOF) regime, where the wind remains slow.
We focus on the long-term formation of circumbinary disks and explore the effects of binary eccentricity and mass ratio.
Our simulations extend to significantly longer timescales, allowing us to investigate the long-term evolution of circumbinary disks.
The problem setup is described in Section~\ref{setup}.
In Section~\ref{results}, we present the simulation results, including the spiral density patterns produced by fast winds and the formation of circumbinary disks through WRLOF under different binary parameters.
Our conclusions are summarized in Section~\ref{sec:conclusion}.

\section{Problem Setup}
\label{setup}

\begin{table*}[t]
    \begin{center}
	\caption{Parameters for simulations in this work.
             From left to right, the quantities are as follows: simulation name, AGB star's mass $M_1$, companion's mass $M_2$, binary mass ratio $q=M_2/M_1$, binary's semi-major asix $a_b$, particle injected velocity $v_{\rm in}$, specific energy ratio $\eta$, binary eccentircity $e_b$, particle injected distribution, and equation of state (EoS). }
	\label{tab:simpara} 
	\begin{tabular}{cccccccccc}
        \toprule
        Name  &  $M_1\,(M_\odot)$  &  $M_2\,(M_\odot)$  &  q  &  $a_b\,(\rm AU)$  &  $v_{\rm in}\, \rm (km/s)$  &    $\eta$    &   $e_b$   &   Distribution   &   EoS \\
        \hline
        run~1   &   $0.73$   &   $0.73$  &  $1.0$  &   $5.0$   &   $42.00$   &   $3.124$   &   0    &    Uniform   &   Locally Isothermal \\ 
        run~2   &   $0.73$   &   $0.73$  &  $1.0$  &   $5.0$   &   $42.00$   &   $3.124$   &   0    &    Gaussian   &   Locally Isothermal \\ 
        run~3   &   $0.73$   &   $1.4$  &  $1.92$  &   $5.0$   &   $39.23$   &   $-0.200$  &   0    &    Uniform   &   Locally Isothermal \\ 
        run~4   &   $0.73$   &   $1.4$  &  $1.92$  &   $5.0$   &   $39.23$   &   $-0.200$  &   0    &    Gaussian   &   Locally Isothermal \\ 
        run~5   &   $0.73$   &   $1.4$  &  $1.92$  &   $5.0$   &   $42.00$   &   $0.781$   &   0    &    Gaussian   &   Locally Isothermal \\ 
        run~6   &   $0.73$   &   $1.4$  &  $1.92$  &   $5.0$   &   $42.00$   &   $0.781$   &   0.2  &    Gaussian   &   Locally Isothermal \\ 
        run~7   &   $0.73$   &   $1.4$  &  $1.92$  &   $5.0$   &   $42.00$   &   $0.781$   &   0.4  &    Gaussian   &   Locally Isothermal \\ 
        run~8   &   $1.063$  &  $1.063$  &  $1.00$  &   $5.0$   &   $46.52$   &     $0.781$   &   0    &    Gaussian   &   Locally Isothermal \\ 
        run~9   &   $1.4$    &   $0.73$  &  $0.52$  &   $5.0$   &   $50.64$   &     $0.781$   &   0    &    Gaussian   &   Locally Isothermal \\ 
        run~10   &   $0.73$   &   $1.4$  &  $1.92$  &   $5.0$   &   $42.00$   &   $0.781$   &   0    &    Gaussian   &   Binary Isothermal \\ 

        \hline
	\end{tabular}
    \end{center}
\end{table*}
We use the SPH code {\sc phantom} (\citealt{2010MNRAS.405.1212L,2010MNRAS.406.1659P,2018PASA...35...31P}) to simulate a binary system consisting of an AGB star and a companion. 
The stars are treated as sink particles. The mass and angular momentum of any gas particles that move inside of their accretion radius are added to the star \citep{Bate1995}.
An AGB star has a large radius of about $1\, \rm AU$.  
Therefore, we set the accretion radius of the AGB star to $R_{\rm s,1} = 1.0 \, \rm AU$, while the companion has an accretion radius of $R_{\rm s,2} = 0.4 \, \rm AU$. 
The semi-major axis of the system is $a_b=5 \, \rm AU$ and the binary eccentricity is $e_{\rm b}$.
The setups for different runs are listed in Table~\ref{tab:simpara}. 
We consider binary masses based on the AC Her system, which hosts a post-AGB star with a mass of $M_1=0.73 \,{\rm M_{\odot}}$ and a main-sequence companion of $M_2=1.4 \,{\rm M_{\odot}}$ (\citealt{2015A&A...578A..40H, 2021A&A...648A..93G, 2023ApJ...950..149A, 2023ApJ...957L..28M}). 
When investigating the impact of the binary mass ratio $q=M_2/M_1$, we modify the masses of the AGB star and the companion while keeping the total binary mass fixed (runs~5, 8, and 9).

For the gas around the binary system, we adopt a local isothermal equation of state with a sound speed following $c_s/v_{\rm Kep}=H/R = 0.1$, where $v_{\rm Kep}$ is the Keplerian velocity, $H$ is the scale height of the disk, and $R$ is the spherical radius from the center of mass of the binary. 
Also note that this is equivalent to $c_s^2 = c_{s,0}^2\cdot R^{-1}$, where $c_{s,0}$ is a constant, since $v_{\rm Kep}\propto R^{-0.5}$. 
We adopted the standard prescription for imposing an effective disc viscosity in {\sc phantom}.
The corresponding Shakura--Sunyaev viscosity (\citealt{1973A&A....24..337S}) parameter can be estimated using Equation~38 of \citet{2010MNRAS.405.1212L},
\begin{equation}
\alpha \approx \frac{1}{10}\alpha^{\rm AV}\frac{\langle h \rangle}{H},
\end{equation}
where $\alpha^{\rm AV}$ is the artificial viscosity parameter in {\sc phantom}, $\langle h \rangle$ is the azimuthally averaged smoothing length, and $H$ is the local disk scale height. 
For the present simulations, this relation gives an effective viscosity of approximately $\alpha \sim 0.1$.

We simulate the AGB star wind by continually injecting gas particles around the AGB star. 
The particles are injected at a radius of $R_1=1.1\, \rm AU$ from the center of the AGB star with a constant radial velocity relative to the AGB star. 
The total injection rate is $\dot{M}=1\times 10^{-6}\,\rm  M_\odot yr^{-1}$. 
With a mass of each gas particle at $2\times10^{-10}\,\rm  M_\odot$, this gives a particle injection rate of $\dot{n} = 5\times10^3 \,\rm yr^{-1}$, which is about $3.8 \times 10^4$ particles per binary orbit for most of our setups. 
However, such an injection rate is not equivalent to the physical mass loss rate for the AGB star because many injected particles will fall back to the AGB star directly due to the viscosity between the gas particles. 

\begin{figure}
    \includegraphics[width=\columnwidth]{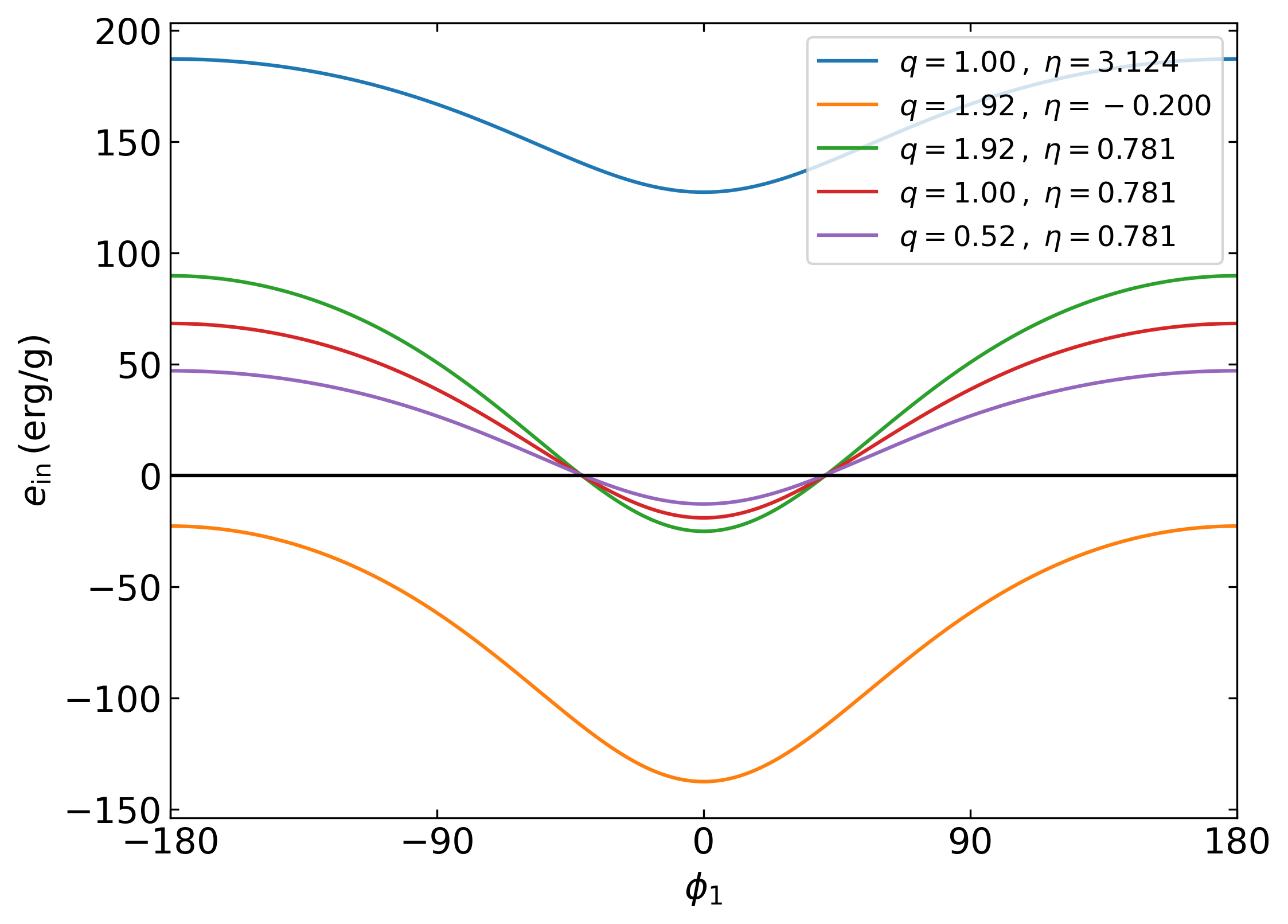}
    \caption{
             Specific energy $e_{\rm in}$ for the injected particles as a function of the angle $\phi_1$, measured relative to the direction toward the companion, for various binary mass ratios $q$ and specific energy ratio $\eta$. The angle $\phi_1=0^\circ$ corresponds to the point on the stellar surface facing the companion.
    }
    \label{fig:SpecificEnergy}
\end{figure}
Although we adopted the binary mass setup based on the AC Her system, our model is scale-free when combined with the injection velocity for the wind. 
Whether a circumbinary disk can form depends entirely on the specific energy of the injected particles $e_{\rm in} = \frac{1}{2}v_{\rm in}^2 + \Phi_1(R_1) + \Phi_2(R_2)$, where $v_{\rm in}$ is the constant injection velocity, $\Phi_1(R_1)$ and $\Phi_2(R_2)$ are the specific potential energy at the inject position relative to the AGB star and the companion. 
The specific energy for the particles injected into the binary orbital plane is shown in Fig.~\ref{fig:SpecificEnergy}. 
A particle with $e_{\rm in }>0$ tends to outflow from the binary system while $e_{\rm in }<0$ tends to be accreted or remain bound. 
Since all injected particles have the same initial distance from the AGB star, $R_1 = 1.1,\mathrm{AU}$, the gravitational potential $\Phi_1(R_1)$ is constant.
However, the distance $R_2$ relative to the companion varies over the AGB star surface. 
Consequently, $\Phi_2(R_2)$ and $e_{\rm in}$ vary for different particles. 
Figure~\ref{fig:SpecificEnergy} shows that the particles injected close to the companion ($\phi_1 = 0^\circ$) have a lower specific energy, which makes it harder for them to escape from the binary potential. 
On the other hand, the particles on the opposite side of the star ($\phi_1=180^\circ$) have a higher specific energy and escape more easily. 

To quantify various injection velocities in different systems, we define a specific energy ratio, 
\begin{equation}
    \eta = \frac{e_{\rm max}}{e_{\rm max}-e_{\rm min}}\,,
    \label{eq:eta}
\end{equation}
where $e_{\rm max}$ and $e_{\rm min}$ are the maximum and minimum specific energy of the injected particles, corresponding to $\phi_1=180^\circ$ and $\phi_1=0^\circ$, respectively, where $\phi_1$ is defined as the angle of the injected particle relative to the direction toward the companion.
For the simulations with $\eta > 1$, all particles have a specific energy $e_{\rm in}>0$ and can escape from the binary system.
For $\eta < 0$, all particles have a specific energy $e_{\rm in}<0$ and eventually are accreted by the binary.
If $0\leq\eta\leq1$, a portion of the particles are accreted, while the others can escape. 
Typically, the gas particles with $e_{\rm in}<0$ may not necessarily be directly accreted. 
They could escape the circumsingle orbit and enter a circumbinary orbit, where they lose angular momentum before being accreted.


In general, gas particles are injected isotropically on the surface of the AGB star. 
However, in disk formation scenarios, we focus on gas near the binary orbital plane, as most of the gas in the polar directions escapes the system without contributing to the disk formation. 
Therefore, we introduce an equatorially biased injection scheme to enhance the resolution near the disk mid-plane. 
Such a preference is defined by a Gaussian-distributed injection rate
\begin{equation} 
    \dot{n}(\mu) = A\exp{\left(-\frac{\mu^2}{2\Delta^2}\right)}, 
\end{equation}
where $A$ is a normalization constant with units of per area per time, $\mu = \cos{\theta}$, $\theta$ is the polar angle on the stellar surface measured from the direction of the binary angular momentum vector (perpendicular to the binary orbital plane), and $\Delta = 0.5$ controls the angular width of the distribution. 
Under this Gaussian-weighted angular distribution, the total particle injection rate over the stellar surface becomes 
\begin{align} 
    \dot{N}_{\rm G} &= 2\pi\int_{-1}^{1}{\;R_{1}^2\cdot A\exp{\left(-\frac{\mu^2}{2\Delta^2}\right)}}d\mu  \\
    &\approx 0.6\cdot4\pi R_{1}^2A = 0.6\cdot\dot{N}_{\rm uniform}, 
\end{align}
where $\dot N_{\rm uniform}$ is the injection rate under a uniform angular distribution. 
As a result, while the injection rate $\dot{M}$ is kept constant in the simulation, the corresponding physical mass-loss rate $\dot{M}_{\rm true}$ is effectively increased by a factor of approximately $1.6$  due to the Gaussian angular weighting.
To ensure the same injected density at the mid-plane, we set a total injection rate of $\dot{M}=1\times10^{-6}\,\rm M_\odot\,yr^{-1}$ for the Gaussian model, and $\dot{M}=1.6\times10^{-6}\,\rm M_\odot\,yr^{-1}$ for the uniform model. 

\section{Results}
\label{results}

We first present results that show the formation of spiral structures and circumbinary disks in simulations with both uniform and Gaussian injection. 
In the following subsections, we examine the dependence of disk formation on the binary eccentricity and mass ratio using the Gaussian injection to study the disk in greater detail.

\subsection{Spiral Structure and Disk Formation}
\begin{figure*}[htb!]
    \centering
    \includegraphics[width=1.0\textwidth]{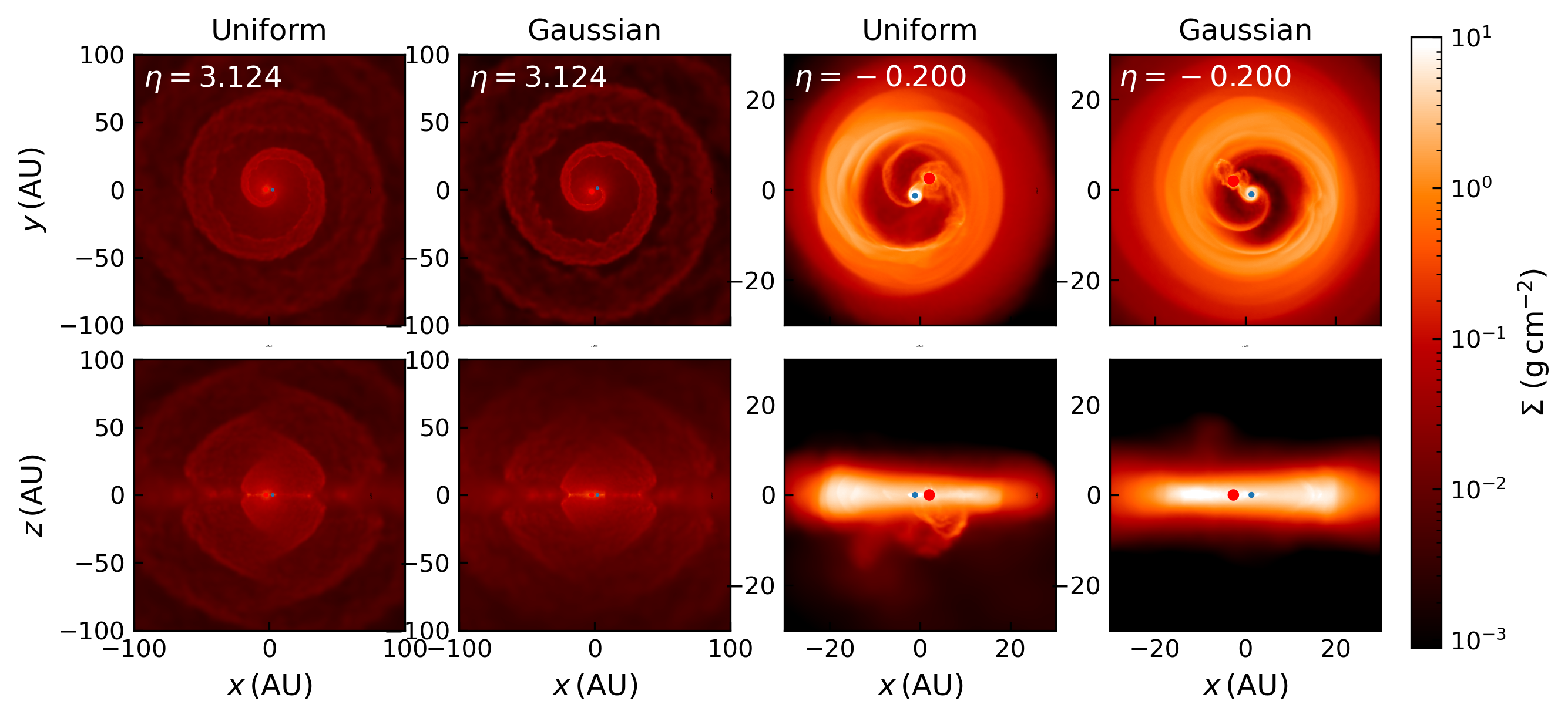}
    \caption{
             Snapshots of the projected density at $600$ orbital periods.
             The top and bottom rows show the density projected along the $z$- and $y$-axes, respectively.
             From left to right, the columns correspond to simulations with $\eta = 3.14$ for Uniform and Gaussian injection (runs 1 and 2; BH accretion) and $\eta = -0.200$ for Uniform and Gaussian injection (runs 3 and 4; WRLOF).
             The red and blue circles mark the AGB star and the companion, respectively, with both shown at their actual accretion radius.
    }
    \label{fig:SpiralAndDisk}
\end{figure*}
Spiral structure has been discovered observationally around AGB stars (\citealt{2006A&A...452..257M,2012Natur.490..232M,2015ApJ...814...61K}), and is also confirmed by simulations (\citealt{1998ApJ...497..303M,2008ApJ...675L.101E,2017MNRAS.468.4465C,2020ApJ...892..110C,2023A&A...674A.122E}). 
In particular, \cite{2017MNRAS.468.4465C} and \cite{2023A&A...674A.122E} performed simulations that include dust formation and radiative transfer processes, which naturally lead to the formation of AGB star winds. 
Their results, therefore, provide a useful benchmark for validating our wind injection model.

Figure~\ref{fig:SpiralAndDisk} shows snapshots for four simulations with  $\eta = 3.124$ and $\eta = -0.200$ for both the uniform and Gaussian wind injection models (runs~1-4). 
For $\eta = 3.124$, all particles are injected with a specific energy much greater than zero, corresponding to a strong wind launched from the AGB star.
Such wind interacts with the companion through the BH accretion. 
As expected, both the uniform and Gaussian injection models exhibit a clear spiral structure, consistent with that reported in \cite{2023A&A...674A.122E}.
We emphasize that this spiral structure is distinct from the spiral arms commonly found in accretion disks.
The gas flow within this spiral structure is instead dominated by radial velocity and therefore the material tends to escape the binary system directly.
This structure arises from the gravitational focusing effect of the companion, which concentrates the wind material toward the mid-plane as the outflow passes near the companion.

\begin{figure}[htb!]
    \centering
    \includegraphics[width=1.0\columnwidth]{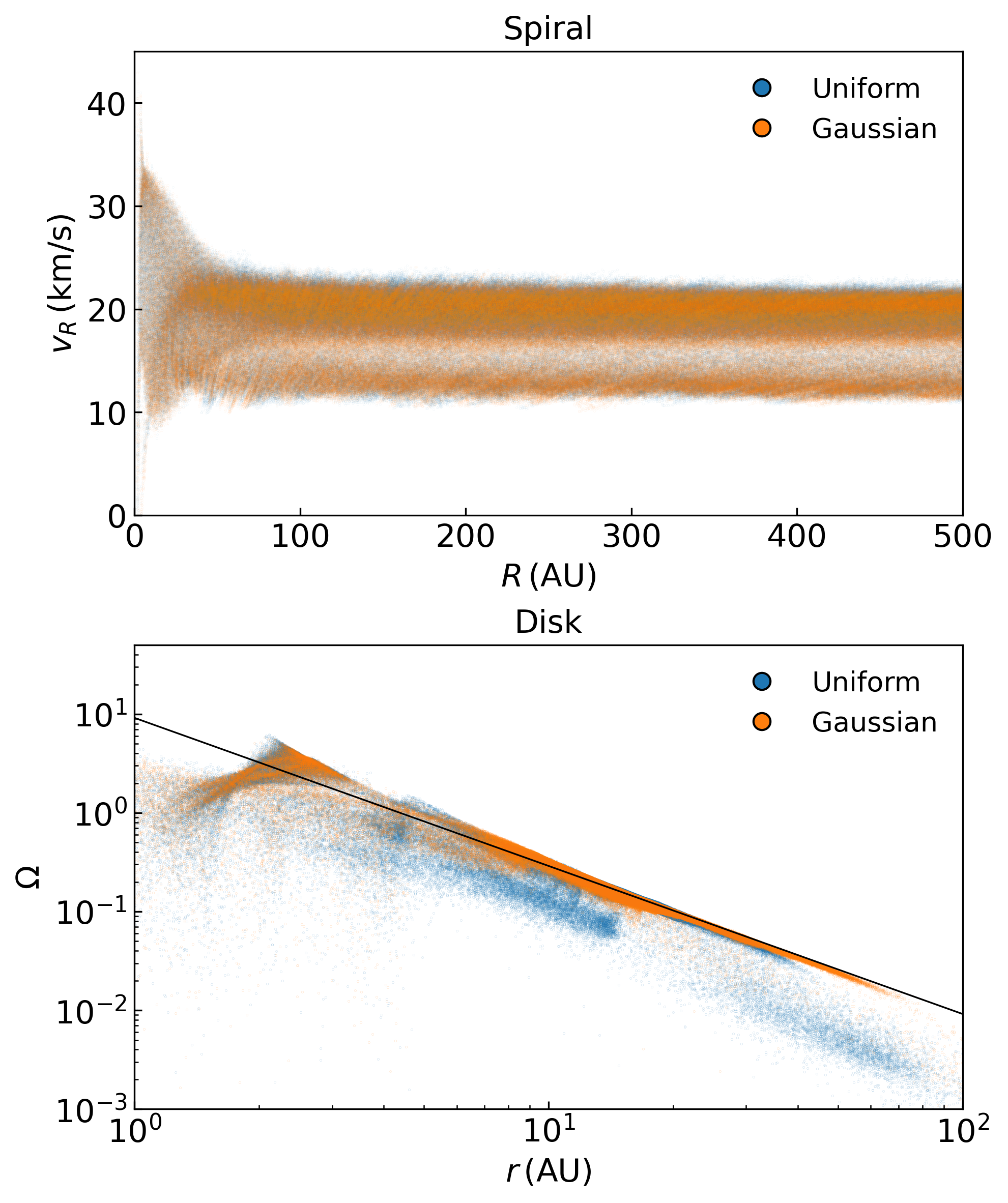}
    \caption{
            Scatter plots of particle velocities in a snapshot at $600$ orbital periods. 
            The top panel shows the particle radial velocity ($v_R$) as a function of the spherical radius ($R$) for runs~1 and 2, which form the spiral structure.
            The bottom panel shows the particle angular frequency ($\Omega$) as a function of the cylindrical radius ($r$) for runs~3 and 4, which form circumbinary disks. 
            Blue and orange dots represent the uniform and Gaussian injection cases, respectively. 
            The black line in the bottom panel shows the Keplerian angular frequency for a central object with mass $M_b = M_1 + M_2$.
            }
    \label{fig:VelSpiral}
\end{figure}

This behavior is more clearly illustrated in the radial velocity map of gas particles shown in the upper panel of Figure~\ref{fig:VelSpiral}.
As the wind propagates outward, it experiences a mild deceleration and ultimately leaves the binary potential with an approximately constant velocity.
Two distinct groups of outflowing gas can be identified: one with a terminal velocity of $\sim 20\,\mathrm{km\,s^{-1}}$ and another at $\sim 14\,\mathrm{km\,s^{-1}}$.
The slower component corresponds to gas that is more strongly decelerated by the companion.
The resulting terminal wind velocities are consistent with those reported in \cite{2023A&A...674A.122E}, where values between $15\,\mathrm{km\,s^{-1}}$ and $25\,\mathrm{km\,s^{-1}}$ are found for different radiative transfer models.

When the wind strength is milder and the companion is more massive, the injected gas has a specific energy below zero, corresponding to $\eta = -0.200$.
In this regime, the wind is slow and fills the AGB star's Roche lobe, leading to WRLOF.
The stronger gravitational potential of the companion is able to capture gas and form a circumsingle disk around the companion, as shown in the last two columns of Figure~\ref{fig:SpiralAndDisk}.
The presence of the circumsingle disk significantly increases the effective cross section of the companion to the outflow and consequently enhances the efficiency of angular momentum transfer.
Once the flow gains sufficient angular momentum, it can escape from the circumsingle orbit and transition to a circumbinary orbit at L2, forming a coherent outflowing stream \citep{Shu1979}.
These materials subsequently accumulate, leading to the formation and outward expansion of a circumbinary disk.
This flow morphology is consistent with the circumbinary outflow predicted in the WRLOF model of \cite{2007ASPC..372..397M} (see also \citealt{2019A&A...626A..68S,2025ApJ...990..172S}), and with the circumbinary disk formation reported in the simulations of \cite{2017MNRAS.468.4465C,2020ApJ...892..110C}, which include radiative transfer and dust formation in AGB star winds.
The existence of the circumbinary disk is further confirmed by the angular frequency map shown in the bottom panel of Figure~\ref{fig:VelSpiral}.
Most gas particles lie along the Keplerian angular frequency curve (black line), starting at $r \simeq 6\,\mathrm{AU}$, for both the uniform and Gaussian injection models.
An additional group of particles located at $r \simeq 2\,\mathrm{AU}$ corresponds to the circumsingle disk around the companion.

For both the spiral-structure and the disk-formation regimes, the differences between the uniform and Gaussian injection models are generally minor.
In the spiral-structure cases, the two models produce nearly identical morphologies, with the only noticeable difference being a lower density at higher latitudes in the Gaussian injection model, as expected from the Gaussian profile in $\mu$.
In the disk-formation regime, although both models inject the same flow density at the mid-plane, the Gaussian injection model produces a denser and radially wider disk than the uniform injection model.
This difference arises because wind material injected at higher latitudes in the Gaussian model is more likely to fall back onto the disk and gain angular momentum through interaction with the disk flow.
As shown in the bottom panel of Figure~\ref{fig:VelSpiral}, in the uniform injection model, a group of gas particles at $r \simeq 10\,\mathrm{AU}$ rotates at a sub-Keplerian angular frequency.
This leads to a higher accretion rate onto the binary and, consequently, a reduced disk expansion rate.

\begin{figure*}[htb!]
    \centering
    \includegraphics[width=1.0\textwidth]{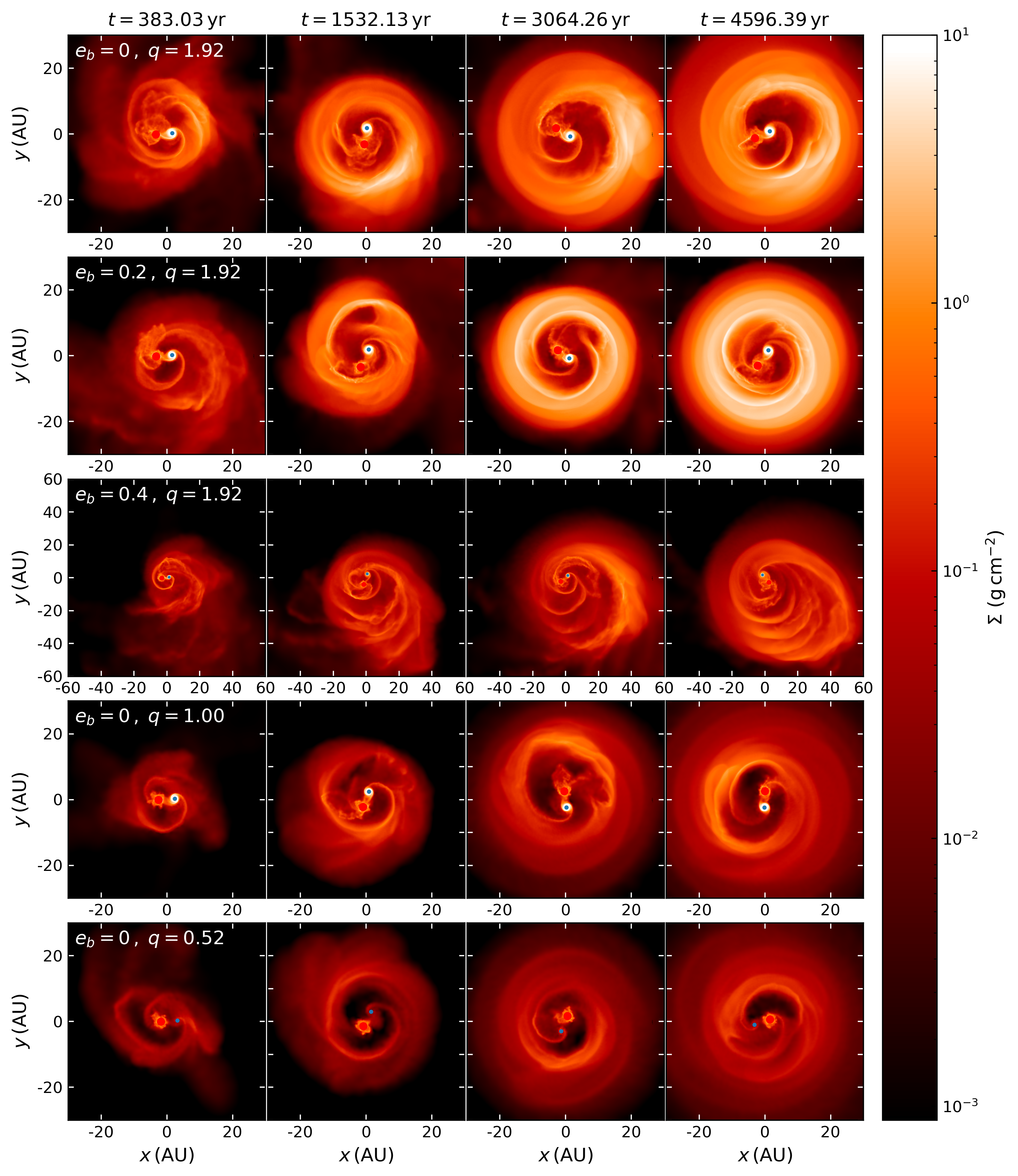}
    \caption{
             Snapshots of projected density for different binary eccentricity $e_b$ and mass ratio $q$. 
             From left to right, the columns correspond to $t = 383.03\,\mathrm{yr}$, $1532.13\,\mathrm{yr}$, $3064.26\,\mathrm{yr}$, and $4596.39\,\mathrm{yr}$, ($50$, $200$, $400$, and $600$ binary orbital periods,) respectively.
             From top to bottom, the rows show the results for models with $(e_b=0,\,q=1.92)$, $(e_b=0.2,\,q=1.92)$, $(e_b=0.4,\,q=1.92)$, $(e_b=0,\,q=1.00)$, and $(e_b=0,\,q=0.52)$, respectively (runs 5-9). 
    }
    \label{fig:DiskEcc}
\end{figure*}

Aside from these secondary differences, the overall disk structures produced by the two injection models are essentially identical.
Therefore, balancing physical realism, computational cost, and numerical resolution, we adopt the Gaussian injection model in the following sections when focusing on disk formation.
To explore a more general regime beyond the extreme spiral-structure and disk-formation cases, we adopt a higher injection velocity corresponding to $\eta = 0.781$.
In this regime ($0 < \eta < 1$), part of the outflow remains gravitationally bound to the binary system, while the rest escapes, likely leading to WRLOF.
We then systematically vary the binary eccentricity, $e_b = 0,\,0.2,$ and $0.4$, as well as the binary mass ratio, $q = 1.92,\,1.0,$ and $0.52$, while keeping the total binary mass fixed.
We run these simulations for over $4500\,\mathrm{yr}$ (over 600 binary orbits) to explore the formation and long-term evolution of the circumbinary disks. 
Time-series snapshots for these models are presented in Figure~\ref{fig:DiskEcc}.

\subsection{Binary Eccentricity}
In this section, we examine the effect of the binary eccentricity on circumbinary disk formation.
The first three rows of Figure~\ref{fig:DiskEcc} correspond to three different binary eccentricities (runs~5-7).

For $e_b = 0$, a dense circumsingle disk forms around the companion at an early stage (first column), while only a small amount of material initially occupies circumbinary orbits, appearing as several spiral arms.
This proto--circumbinary disk, however, acts as an efficient barrier to the AGB star wind.
As the outflow crosses the proto-disk, shocks are generated, angular momentum is exchanged, and the incoming material merges with the rotating flow.
Through this process, the circumbinary disk becomes progressively denser and expands radially.
For $e_b = 0.2$, the disk formation process is broadly similar to that in the $e_b = 0$ case.
However, the resulting circumbinary disk is narrower and more compact, and two well-defined spiral arms are present within the disk.
For $e_b = 0.4$, the flow morphology changes substantially.
Although a circumsingle disk still forms around the companion, the highly eccentric binary orbit allows it to approach much closer to the AGB star.
Near periastron, the L2 point moves closer to the companion, enabling a larger fraction of material from the inner disk to escape at high velocity and form a prominent large-scale spiral arm.
This outflow is injected into circumbinary orbits, where it follows a highly eccentric trajectory and ultimately gives rise to an eccentric circumbinary disk.
\begin{figure*}[htb!]
    \centering
    \includegraphics[width=1.0\textwidth]{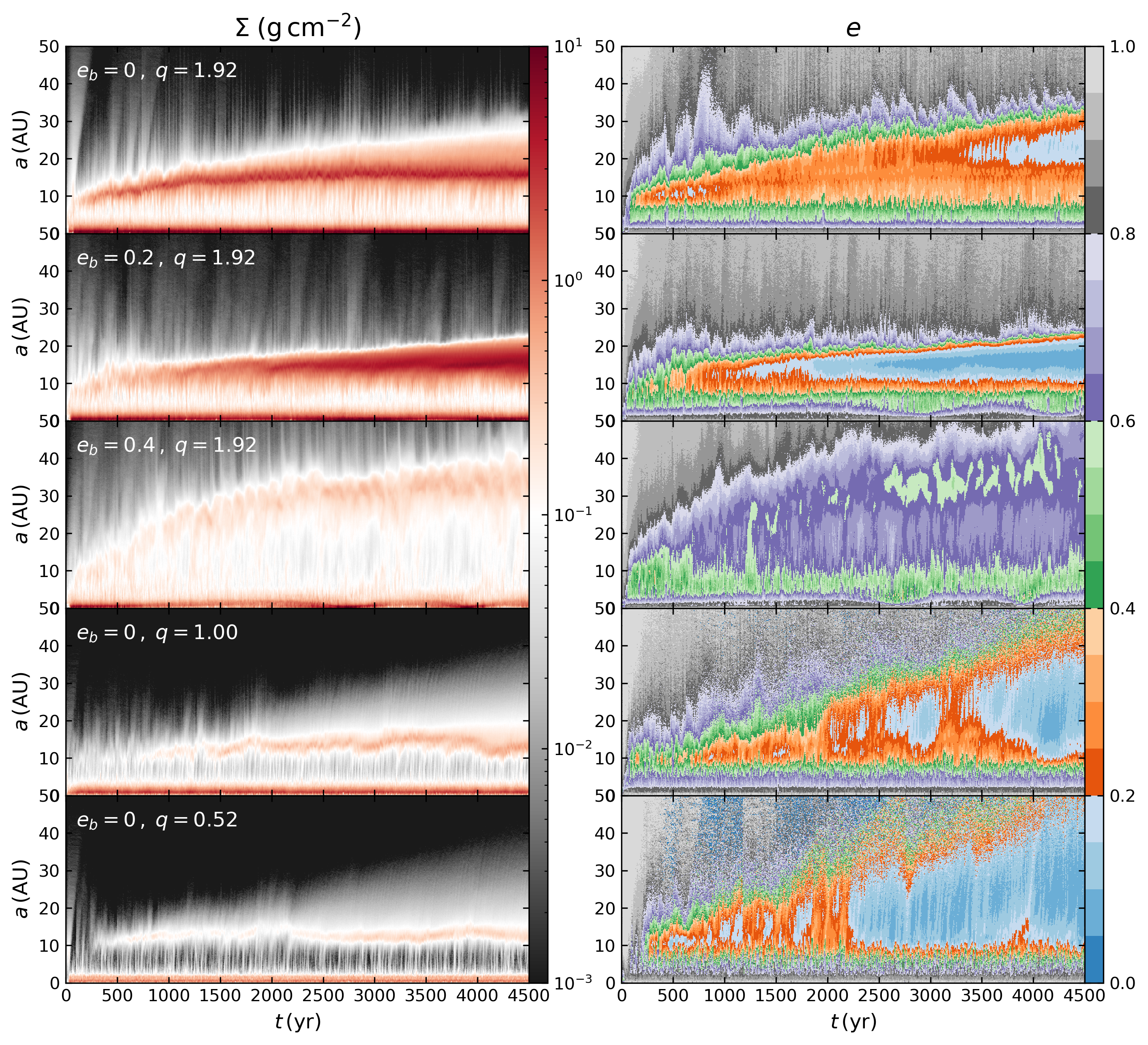}
    \caption{
             Surface density and eccentricity evolution for the circumbinary disk. 
             Here $e_{\rm d}$ represents the mass-averaged particle eccentricity magnitude and is used as a statistical measure of orbital non-circularity.
             From top to bottom, the rows show the results for models with $(e_b=0,\,q=1.92)$, $(e_b=0.2,\,q=1.92)$, $(e_b=0.4,\,q=1.92)$, $(e_b=0,\,q=1.00)$, and $(e_b=0,\,q=0.52)$, respectively (runs 5-9). 
    }
    \label{fig:EccVsA}
\end{figure*}

To better characterize the disk structure, we perform a statistical analysis of the gas particles.
For each particle, we first compute the semi-major axis $a_p$ and eccentricity $e_p$, assuming motion in a Keplerian potential.
The semi-major axis is calculated as
\begin{equation}
    a_p = -\frac{GM_b}{2} \cdot \left[\frac12 \mathbf{v}_p^2 - \frac{GM_b}{\left|\mathbf{R}\right|}\right]^{-1}\,,
\end{equation}
and the eccentricity is given by 
\begin{equation}
    e_p = \left|{\frac{\mathbf{v}_p\times\boldsymbol{\ell}_p}{GM_b} - \frac{\mathbf{R}}{\left|\mathbf{R}\right|}}\right|\,,
\end{equation}
where $M_b = M_1 + M_2$ is the total binary mass, $\mathbf{v}_p$ is the particle velocity relative to the binary center of mass, $\mathbf{R}$ is the position vector relative to the binary center of mass, and $\boldsymbol{\ell}_p = \mathbf{R} \times \mathbf{v}_p$ is the particle specific angular momentum.
We then construct a one-dimensional grid in semi-major axis $a$ and compute the surface density
\begin{equation}
    \Sigma(a) = \frac{\sum_n m_p}{2\pi a\, \mathrm{d}a}\,,
\end{equation}
as well as the disk-averaged eccentricity
\begin{equation}
    e_{\rm d}(a) = \frac{\sum_n e_p}{n}\,,
\end{equation}
within each annulus, where $n$ is the number of particles in a radial bin of width $\mathrm{d}a = 0.2\,\mathrm{AU}$.
The temporal evolution of the disk surface density and eccentricity is shown in Figure~\ref{fig:EccVsA}.

The first three rows of Figure~\ref{fig:EccVsA} illustrate the disk structure for different binary eccentricities.
Clear differences emerge between the $e_b = 0$ and $e_b = 0.2$ cases.
For $e_b = 0$, the peak surface density of the disk remains nearly constant, while the disk continues to expand radially from $\sim 10\,\mathrm{AU}$ to $\sim 30\,\mathrm{AU}$.
During this process, the disk maintains a moderate eccentricity of $e_{\rm d} \simeq 0.2$ and gradually circularizes at late times.
In contrast, for $e_b = 0.2$, the disk is significantly narrower, confined between $\sim 10\,\mathrm{AU}$ and $\sim 20\,\mathrm{AU}$, expands more slowly, and becomes nearly circular at an earlier stage.

For the highly eccentric case with $e_b = 0.4$, as discussed above, a substantial fraction of the flow is injected directly into highly eccentric circumbinary orbits.
Consequently, the disk mass is concentrated at larger semi-major axes around $a \simeq 30\,\mathrm{AU}$, with a characteristic eccentricity of $e_{\rm d} \simeq 0.6$.

Different from the $e_b=0.4$ case, for $e_b = 0$ and $0.2$, the disk eccentricity is generated through binary--disk interactions after the disk has formed.
While viscous dissipation in a dense disk tends to circularize the flow, gravitational torques from the binary can excite non-circular motion, even for circular binaries (\citealt{2020ApJ...889..114M,2022MNRAS.516.5446L}).
The competition between these two processes determines the resulting disk eccentricity.
In particular, the $e_b = 0.2$ case exhibits a more rapid circularization than the $e_b = 0$ case, leading to a narrower and more axisymmetric disk structure.

\subsection{Binary Mass Ratio}
The circumbinary disk's angular momentum is entirely supplied by the companion.
In this section, we fix the total binary mass and decrease the mass ratio $q$ (runs~5, 8, and 9), which corresponds to a lighter companion and a denser AGB star.
As the companion becomes less massive, its ability to capture the wind is reduced, leading to a slower disk formation process.
At the same time, the deeper gravitational potential of a denser AGB star makes it more difficult for the wind to escape.

As shown in the first and last two rows of Figures~\ref{fig:DiskEcc} and~\ref{fig:EccVsA}, the resulting disk becomes progressively less dense as $q$ decreases.
In the lowest-mass-ratio case ($q=0.52$), no circumsingle disk around the companion is resolved in the simulation, given an accretion radius of $0.4\,\mathrm{AU}$.
Nevertheless, a faint spiral arm is still launched from the companion and continues to supply material to the proto-disk through the L2 point.

From Figure~\ref{fig:EccVsA}, we further find that, although the overall disk density is lower, the disk in the $q = 0.52$ case still extends to a radius of $a \simeq 40,\mathrm{AU}$ at $t = 4500\,\rm yr$, comparable to the $q = 1$ case and wider than the disk formed for $q = 1.92$.
Moreover, disks formed at lower mass ratios are more circular.
Observationally, the post-AGB star system AC~Her is known to host a low-density yet radially extended disk, reaching sizes of up to $\sim 10^3\,\mathrm{AU}$ (\citealt{2023ApJ...950..149A,2023ApJ...957L..28M,2025ApJ...985...65H}).
Our results indicate that an extended disk does not necessarily require an initial high-density phase, but may instead develop directly from low-density material.

\section{Discussion}\label{sec:discussion}
Our results show that a dense circumbinary disk can form from an AGB wind even within a highly simplified model. 
However, although the overall mechanism is expected to remain similar, a number of additional physical processes may influence the detailed evolution of the system.

For the wind model, we simplify the wind-launching process by directly injecting supersonic particles around the AGB star. 
In reality, wind acceleration is a complex process that is closely linked to dust formation (\citealt{2008A&A...491L...1H,2014MNRAS.442.1440S}), making both thermodynamics (\citealt{2020ApJ...892..110C,2024A&A...691A..57M}) and chemistry (\citealt{1999A&A...347..594G,1990A&A...235..345G,2006A&A...447..553F}) important. 

Among these effects, thermodynamics is expected to be particularly important because it influences not only the dust formation, but also whether wind material can remain gravitationally bound to the binary.
For example, \citet{2024A&A...691A..84M} found that no accretion disk forms around the companion when H-I cooling is neglected, because excessive post-shock heating prevents the gas from remaining bound. 
Efficient cooling can reduce the total energy of the gas, making it easier for material to stay gravitationally bound and producing an effect analogous to increasing the depth of the binary potential well. 
In our simulations, we adopted a locally isothermal equation of state in which the temperature is prescribed as a fixed function of radius. 
This treatment effectively assumes that thermal energy is removed efficiently enough to maintain the imposed temperature profile, while neglecting explicit heating and cooling processes. As a result, it likely favors the formation of both circumsecondary and circumbinary disks.

\begin{figure}[htb!]
    \centering
    \includegraphics[width=1.0\columnwidth]{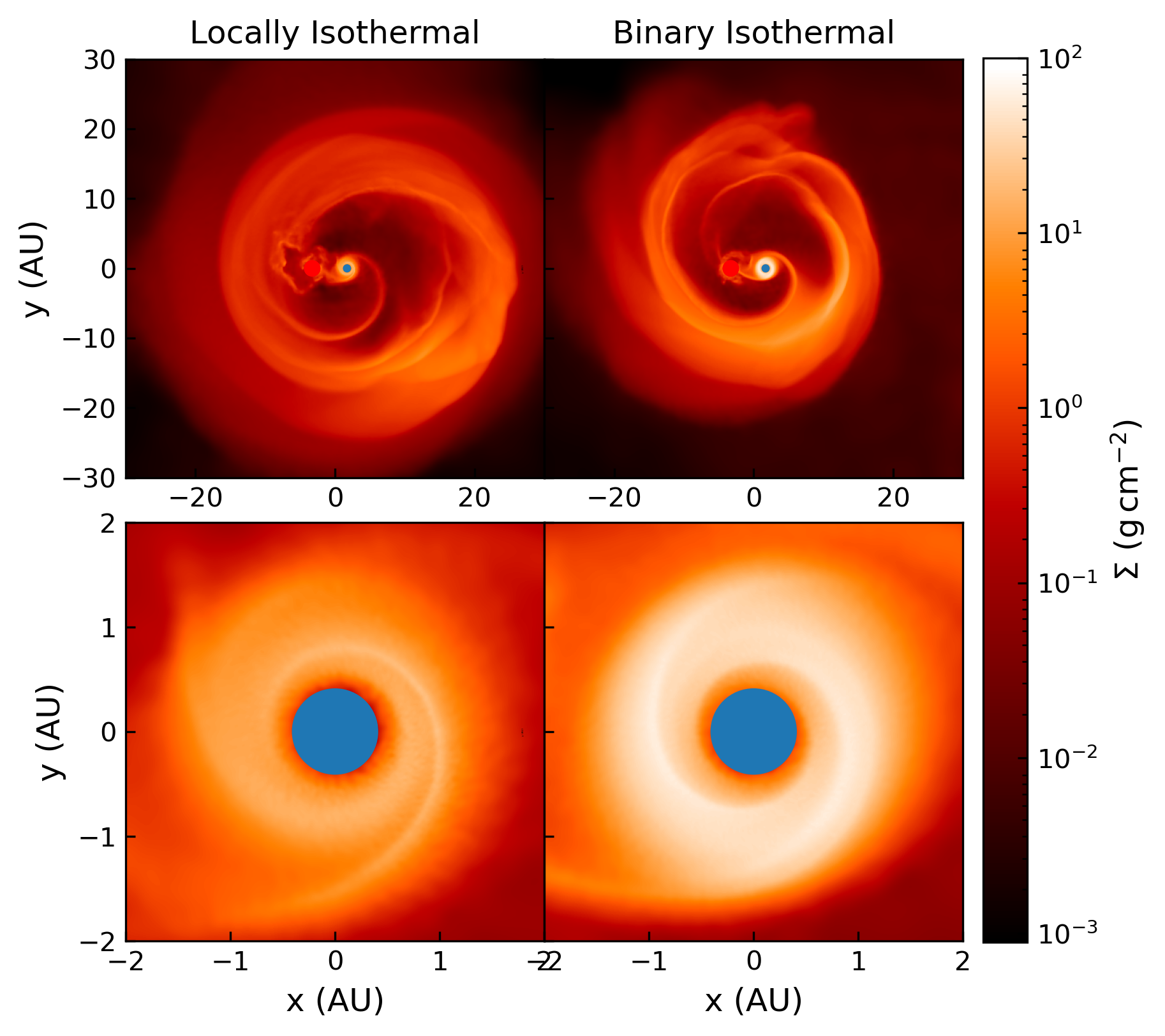}
    \caption{
             Snapshots of the projected density at $t = 2681.22\,\mathrm{yr}$ ($350$ binary orbital periods) for simulations employing different equations of state. 
             The left and right columns show the locally isothermal (run~5) and binary isothermal (run~10) models, respectively. 
             The bottom row presents zoomed-in views of the companion region. 
             The blue circle indicates the companion's accretion radius ($0.4\,\mathrm{AU}$).
    }
    \label{fig:faris}
\end{figure}
Beyond the cooling prescription, the adopted thermodynamic treatment also affects the temperature structure within the system.
The locally isothermal equation of state adopted in this work treats the binary as a single point source and prescribes a temperature profile that decreases with distance from the system center. 
This approximation is reasonable for structures located far from the binary, such as the circumbinary disk. 
However, near the companion it produces an asymmetric temperature distribution, with higher temperatures on the side facing the AGB star and lower temperatures on the opposite side. 
A more realistic treatment considers both stars as independent heating sources and constructs a global temperature profile accordingly (\citealt{2014ApJ...783..134F}).
To evaluate the impact of this assumption, we performed an additional simulation (run~10) using a binary isothermal equation of state, while keeping all other parameters identical to those of run~5. 
The results are shown in Figure~\ref{fig:faris}.
We find that the circumsecondary disk is denser in the binary isothermal model. 
Nevertheless, this difference does not alter the main conclusions of this work, since a similar circumbinary disk forms in both simulations, indicating that the global circumbinary disk structure is relatively insensitive to the adopted temperature prescription.

Furthermore, we adopted a disk aspect ratio of $H/R = 0.1$ in our simulations. 
Observational studies of post-AGB circumbinary disks 
suggest this value could be larger.
For example, \citet{2021A&A...648A..93G} derived aspect ratios ranging from $0.17$ to $0.3$ for several post-AGB systems based on disk-size fitting. 
Similarly, \citet{2024A&A...691A..84M} found that $H/R$ varies between approximately $0.1$ and $0.3$ in simulations including H-I cooling (see also \citealt{2016A&A...596A..92K,2023MNRAS.524.4168A}). 
A larger aspect ratio implies a higher gas temperature and therefore a larger thermal energy content.
As a result, the wind material is less easily captured by the binary potential and disk formation becomes less efficient. 
Consequently, the relatively small value of $H/R=0.1$ adopted in this work may favor the formation of circumsecondary and circumbinary disks.

In addition to these physical assumptions, numerical effects may also influence the resulting disk structure.
For instance, resolution is an important factor affecting disk formation. 
In our simulations, the resolution is determined by the number of particles injected per unit time. 
If the resolution is too low, the circumsecondary disk may not be adequately resolved, and only the large-scale spiral structures can be identified. 
In contrast, our simulations achieve sufficient resolution to resolve both the circumsecondary and circumbinary disks. 
For example, runs~5 and 6 contain more than $10^6$ particles after $600$ binary orbits, allowing the detailed disk morphology to be captured reliably. 

Another potential numerical concern is the role of artificial viscosity in SPH simulations.
For the adopted effective viscosity parameter $\alpha=0.1$, to evaluate whether the imposed effective viscosity significantly influences the disk evolution, we estimated the angular momentum transport associated with the large-scale non-axisymmetric structures in the circumbinary disk. 
We define a local stress proxy,
\begin{equation}
\alpha_{\rm eff} \equiv \frac{v_r(v_\phi-v_{\rm Kep})}{c_s^2},
\end{equation}
where $v_r$ and $v_\phi$ are the radial and azimuthal velocities of an SPH particle, respectively, and $v_{\rm Kep}$ is the local Keplerian velocity.
This quantity characterizes the transport driven by spiral waves, shocks, and radial flows. 
For the circularized disk (run~6), within the circumbinary disk, $\alpha_{\rm eff}$ exhibits substantial spatial fluctuations driven by spiral arms and shocks, with typical values ranging between approximately $-0.3$ and $0.3$, while larger amplitudes are found near the wind--disk interface.
For the eccentric disks, $\alpha_{\rm eff}$ is even larger. 
These stresses are therefore comparable to or greater than the estimated imposed effective viscosity level, indicating that angular momentum transport is primarily governed by physical disk structures rather than the imposed effective viscosity. 
Consequently, the imposed effective viscosity is unlikely to dominate the formation and evolution of the circumbinary disk in our simulations.

\section{conclusions}
\label{sec:conclusion}

In this work, we investigate the formation and dynamical properties of circumbinary disks around AGB star binaries through SPH simulations.
We find that the interaction between the AGB star wind and the companion gives rise to either large-scale spiral structures or circumbinary disks, depending on the binary orbital parameters.
When the wind is relatively fast or the companion is insufficiently massive, corresponding to a Bondi-Hoyle accretion regime, the outflow remains largely unbound and forms a persistent spiral structure shaped by the orbital motion.
In contrast, for milder winds and more massive companions, corresponding to the WRLOF regime, a circumsingle disk can form around the companion, which significantly enhances angular momentum transfer and enables material to be transported through the L2 point into circumbinary orbits, leading to the formation and long-term growth of a circumbinary disk. 

We perform simulations over $\sim600$ binary orbital periods, significantly longer than in previous studies, allowing us to investigate the long-term evolution of circumbinary disks.
The resulting disks exhibit a wide range of radial extents and eccentricities, depending sensitively on the binary orbital parameters.
For circular and mildly eccentric binaries ($e_b = 0$ and $0.2$), the disks form at relatively small radii and gradually expand outward, while maintaining moderate orbital non-circularity.
In contrast, for highly eccentric binaries ($e_b = 0.4$), a significant fraction of the outflow is injected directly into highly eccentric circumbinary orbits, leading to a disk that is concentrated at larger semi-major axes and retains a high eccentricity.

We further find that decreasing the companion mass, through a lower binary mass ratio, leads to lower circumbinary disk surface densities and slower disk growth.
Nevertheless, even in low-mass-ratio cases where the circumsingle disk is not resolved in the simulation, an extended and nearly circular circumbinary disk can still form.
This suggests that the formation of a radially wide circumbinary disk does not necessarily require an initially dense or compact disk phase.

Overall, our results demonstrate that binary-wind interactions, particularly in the WRLOF regime, can naturally account for the formation of a wide range of circumbinary disk morphologies observed in AGB and post-AGB star systems.
In this framework, the angular momentum of the circumbinary disk is entirely supplied by the companion, naturally leading to co-planar disk formation.
However, the post-AGB star system AC~Her has been observed to host a widely extended circumbinary disk that is polar-aligned with respect to the binary orbit (\citealt{2015A&A...578A..40H,2023ApJ...950..149A,2023ApJ...957L..28M,2025ApJ...985...65H}).
The origin of the polar angular momentum and the physical mechanism responsible for such polar-aligned circumbinary disks, therefore, remain open questions.

\software{\texttt{PHANTOM} \citep{2018PASA...35...31P}
          }



\bibliography{wind}
\bibliographystyle{aasjournalv7}



\end{CJK*}
\end{document}